\documentclass[twocolumn,showpacs,preprintnumbers,amsmath,amssymb]{revtex4}

\usepackage{graphicx}
\usepackage{dcolumn}
\usepackage{bm}

\begin{document}

\title{An all fiber source of frequency entangled photon pairs}

\author{Xiaoying Li$^{\text{1}}$}\email{xiaoyingli@tju.edu.cn} \author{Lei Yang$^{\text{1}}$, Xiaoxin Ma$^{\text{1}}$, Liang Cui$^{\text{1}}$} \author{ Zhe Yu Ou$^{\text{2}}$}\email{zou@iupui.edu} \author{Daoyin Yu$^{\text{1}}$}

\affiliation{$^{\text{1}}$ College of Precision Instrument and
Opto-electronics Engineering, Tianjin University, \\Key Laboratory of Optoelectronics Information Science and Technology, Ministry of Education, Tianjin, 300072,
P. R. China\\$^{\text{2}}$ Department of Physics, Indiana University-Purdue University Indianapolis, Indianapolis, IN, 46202, USA}

\date{\today}

\begin{abstract}
 We present an all fiber source of frequency entangled photon
pairs by using four wave mixing in a Sagnac fiber loop. Special care is taken to suppress the impurity of the frequency entanglement by cooling the fiber and by matching the polarization modes of the photon pairs counter-propagating in the fiber loop. Coincidence detection of signal and idler photons, which are created in pair and in different spatial modes of the fiber loop, shows the quantum interference in the form of spatial beating, while the single counts of the individual signal (idler) photons keep constant. When the production rate of photon pairs is about 0.013 pairs/pulse, the envelope of the quantum interference reveals a visibility of $(95\pm 2)\%$, which is close to the calculated theoretical limit $97.4\%$.
\end{abstract}

\maketitle

\section{Introduction}

Entangled photon pairs are not only at
the heart of the most fundamental tests of quantum
mechanics, but also essential resources for quantum information processing (QIP). In the past three decades, parametric process in nonlinear media has been proved to be an efficient method to produce entanglement. Spontaneous parametric down-conversion (SPDC) in $\chi ^{(2)}$
nonlinear crystals has been extensively studied, and photon pairs entangled in various degrees of freedom, such as polarization, momentum, frequency and time-energy etc, had been realized. The various degrees of freedom provide more flexibility for studying the fundamental physics and for the applications in quantum information~\cite{Tittel01}. Recently, there has been growing interest
in generating entangled photon pairs by using $\chi ^{(3)}$ nonlinearity in optical fibers~\cite{Li05a,Fan05a,Alibart06,takesue06opex,Chen08pra}. In this situation, entangled photons are produced in well defined fiber modes, which is convenient for integrating with waveguide devices and holds the promise of developing scalable quantum optical devices. However, comparing with its $\chi ^{(2)}$-crystal counterparts, the degrees of freedom of entanglement in optical fiber have not been fully explored.

It has been proved that spontaneous four-wave mixing (SFWM) in optical fibers is an excellent source of quantum-correlated photon
pairs~\cite{Li05a,Fan05a,Alibart06}. In this process, two pump
photons at frequencies $\omega_{p1}$ and $\omega_{p2}$ are scattered through the $\chi ^{(3)}$ (Kerr)
nonlinearity of the fiber to simultaneously create a signal photon
and an idler photon at frequencies $\omega_{s}$ and $\omega_{i}$,
respectively, such that $\omega_s + \omega_i = \omega_{p1} + \omega_{p2}$. When phase matching condition is satisfied, the possibility of SFWM is greatly enhanced. In a dispersion shifted fiber (DSF), SFWM  process is phase-matched for one pulsed pump with the central wavelength close to the zero dispersion wavelength (ZDW) of DSF, or for two pulsed pump with the average of the central wavelengths close to the ZDW. In the former case, we have $ \omega_{p1} = \omega_{p2}$ while in the latter, $ \omega_{p1} \neq \omega_{p2}$.  Because of the isotropic nature of the Kerr nonlinearity in
fused-silica-glass fiber, the generated photon pairs are
predominantly co-polarized with the pump photons. In addition, different kinds of entanglement photon pairs, such as polarization entanglement, path entanglement, and time-bin entanglement etc, can be realized by coherently superposing two SFWM processes~\cite{Li05a,LeeQELS05,takesue06opex,Chen07}.

Frequency entangled two-photon state is made of two photons of two different frequencies. It has the simple single-mode form of
\begin{eqnarray}
\left| \Psi _{f}\right\rangle = {1\over \sqrt{2}}\Big(\left| \omega
_1\right\rangle _c\left| \omega _2\right\rangle
_d+\left| \omega _2\right\rangle _c\left|
\omega _1\right\rangle _d\Big),
\label{f}
\end{eqnarray}
where $\omega_1\ne \omega_2$ and $c,d$ denote different modes. This state was first realized in parametric down-conversion in both type-I \cite{ou88b} and type-II \cite{shih94} forms.
Compared to other types of entanglement, frequency entangled two-photon state is less studied perhaps because it involves photons of different frequencies. Nevertheless, it provides another degree of freedom for quantum information.

In this paper, we experimentally demonstrate the first, to the best of our knowledge, all fiber source of frequency entangled photon pairs by using the spontaneous four-wave mixing (SFWM) in a Sagnac fiber loop (SFL). The background of the frequency entanglement is suppressed not only by cooling the fiber, but also by carefully matching the polarization modes of the photon pairs counter-propagating in the SFL. The nature of the frequency entanglement is evidenced by the coincidence detection of the nondegenerate signal and idler photon
pairs, which shows the quantum interference in the form of spatial beating, while the single count rates of the individual signal (idler) photons are constant. For the photon pairs with a production rate of about 0.013 pairs/pulse, the visibility of interference is $(95\pm 2)\%$ when only the dark counts of detectors are subtracted.

The paper is organized as follows. After discussing the principle of the experiment, we will devote Sect.III to the problem of polarization mode matching, which is essential in realizing high fidelity frequency entanglement. In Sect.IV, we will describe the experimental procedure and analyze and interpret the experimental data. We end with a summary.

\section{Principle of the experiment}

In SFWM in a fiber, two non-degenerate frequency components $\omega_s,\omega_i$ are produced that are in a two-photon state entangled with vacuum \cite{ou89,wang01,chen05}:
\begin{eqnarray}
\left| \Psi _{SFWM}\right\rangle =\left| 0\right\rangle +\eta \left| \omega
_s,\omega _i\right\rangle ,\label{SFWM}
\end{eqnarray}
where $\eta (|\eta|<<1)$ is related to the Kerr nonlinearity of the fiber and the square of the amplitude $E_p$ of the pump field: $\eta\propto E_p^2$. However, the photons are not frequency entangled in the above state. Frequency entanglement is obtained by coherently adding up two counter-propagating SFWM processes in a Sagnac fiber loop (SFL). The principle is similar to that in Ref.~\cite{Chen07}, however, instead of investigating degenerate photon pairs~\cite{Chen07} by using two pulsed pumps with non-degenerate frequencies, we exploit nondegenerate photon pairs by using one pulsed pump.

As illustrated in Fig. 1(a), a SFL consists of a piece of fiber, a fiber polarization controller (FPC), and a 50/50 fiber splitter/coupler. It is preceded by a
circulator (Circ), which redirects the SFL reflected photons to a separate spatial mode. For clarity and consistency, we label the two output modes of SFL in Fig.1(a) as "$c$" and "$d$". The pump injected into the SFL is split into two pumps traversing
in a counter-propagating manner by the 50/50 fiber splitter/coupler. Because of the nature of the 50/50 fiber splitter/coupler, there is a $\pi/2$ phase shift in the cross-coupled field. Each pump then
coherently produces copolarized SFWM photon pairs, $\left| \omega
_s,\omega _i\right\rangle_a$ and $\left| \omega
_s,\omega _i\right\rangle_b$, where the footnotes $a$ and $b$ denote the photon pairs transmitted in the clockwise and counter-clockwise directions. The photon pairs in two directions, with a phase difference controlled by
the FPC in SFL, are then recombined at the coupler
before coming out of SFL. Thus,  the state of the photons before they are recombined in the 50/50 fiber coupler is written as, according to Eq.(\ref{SFWM}),
\begin{eqnarray}
\left| \Psi _{in}\right\rangle =(\left| 0\right\rangle _a+\eta \left| \omega
_s,\omega _i\right\rangle _a)\otimes (\left| 0\right\rangle _b-e^{i2\phi
}\eta \left| \omega _s,\omega _i\right\rangle _b).
\end{eqnarray}
The minus sign in the photon state of the $b$ field is due to the $\pi/2$ phase suffered by the pump field coupled across the  50/50 fiber splitter/coupler (recall $\eta\propto E_p^2$) and $\phi$ is the phase difference between the two counter-propagating pump fields in SFL. The 50/50 fiber splitter/coupler now behaves like an ideal 50/50 symmetric beam splitter to recombine the $a,b$ fields for the output fields $c$ and $d$. It makes the following transformation for the states:
\begin{eqnarray}
\left| \omega\right\rangle_a \rightarrow (\left| \omega\right\rangle _d+ i \left| \omega\right\rangle _c)/\sqrt{2},\\
\left| \omega\right\rangle_b \rightarrow (\left| \omega\right\rangle _c+ i \left| \omega\right\rangle _d)/\sqrt{2}.
\end{eqnarray}
Under the assumption that the modes of the photon pairs $\left| \omega
_s,\omega _i\right\rangle_a$ and $\left| \omega
_s,\omega _i\right\rangle_b$ are ideally matched, the output state of the 50/50 fiber coupler can be expressed as
\begin{eqnarray}
\left| \Psi _{out}\right\rangle =|0\rangle + \sqrt{2}\eta e^{i\phi} |\psi\rangle ,
\label{eq6}
\end{eqnarray}
with
\begin{eqnarray}
\left| \psi\right\rangle ={1\over \sqrt{2}} \Big[\cos\phi(\left| \omega
_s,\omega _i\right\rangle _d-\left| \omega _s,\omega _i\right\rangle
_c) \cr +\sin\phi (\left| \omega _s\right\rangle _c\left|
\omega _i\right\rangle _d+\left| \omega _i\right\rangle _c\left| \omega
_s\right\rangle _d)\Big] ,
\label{eq7}
\end{eqnarray}
where higher order terms in $\eta$ are omitted. From Eq.(\ref{eq7}), we then obtain the state $|\psi\rangle$ as
\begin{eqnarray}
\left| \psi_1\right\rangle =(\left| \omega _s,\omega _i\right\rangle
_c-\left| \omega _s,\omega _i\right\rangle _d)/\sqrt{2}\label{psi1}
\end{eqnarray}
for $\phi =0$ and
\begin{eqnarray}
\left| \psi_2\right\rangle =(\left| \omega _s\right\rangle _c\left|
\omega _i\right\rangle _d+\left| \omega _i\right\rangle _c\left| \omega
_s\right\rangle _d)/\sqrt{2}\label{psi2}
\end{eqnarray}
for $\phi =\pi/2$. In the former case ($\phi =0$), both signal and idler photons of a pair via SFWM come out in the same spatial mode, they are both simultaneously in either mode $c$ or mode $d$~\cite{Fiorentino02,Li05a}. In the latter case ($\phi =\pi/2$), the simultaneously created signal and idler photons have different spatial modes, i.e., if
one photon in mode $c$ is known to be in a frequency $\omega _i$ then the other
one in mode $d$ is determined to have frequency $\omega _s$, or vice versa.  For the signal and idler photons  emerging in different spatial modes, there is no way to determine which pump in SFL created the photon, this indistinguishability gives rise to frequency entanglement, as shown in Eq.(\ref{psi2}), which  has exactly the same form as  Eq.(\ref{f}) for frequency entangled two-photon state.

The confirmation of the two-photon frequency entanglement is similar to that in other degrees of freedom, that is, the observation of quantum interference in two-photon coincidence. For frequency, the corresponding variable is time. Indeed, let us make a time-resolved two-photon coincidence measurement between $c$ and $d$ fields, that is, to measure
\begin{eqnarray}
P_2(\tau) \propto \langle \hat E_d^{(-)}(t)\hat E_c^{(-)}(t+\tau)\hat E_c^{(+)}(t+\tau)\hat E_d^{(+)}(t)\rangle\label{Pt2}
\end{eqnarray}
with
\begin{eqnarray}
\hat E_k^{(+)}(t)= \hat a_k(\omega_s)e^{-i\omega_st} + \hat a_k(\omega_i)e^{-i\omega_it}\cr
[\hat E_k^{(-)}(t)]^{\dag}=\hat E_k^{(+)}(t).~~~~(k=c,d)\label{Ecd}
\end{eqnarray}
Here the quantum average is over the frequency entangled state in Eq.(\ref{psi2}) and we only consider two frequency modes at $\omega_{s,i}$ for the field operators. A straightforward calculation gives
\begin{eqnarray}
P_2(\tau) \propto 1+ \cos(\omega_s-\omega_i)\tau, \label{bt}
\end{eqnarray}
which shows a beat in frequency difference $\omega_s-\omega_i$. So the two-photon quantum interference effect for a frequency entangled two-photon state manifests itself in the form of a temporal beat in time-resolved two-photon coincidence measurement.

On the other hand, for the signal and idler photons that are generated in SFWM in fiber, $\omega_s-\omega_i \sim 10^{13} rad/s$, which requires detectors with a time resolution of 100 fs. The best detectors in current technology have a time resolution of the order of a few ps. So it is impossible to directly observe the temporal beat for a verification of frequency entanglement. However, we may employ some interference methods to explore the quantum superposition in Eq.(\ref{psi2}). The simplest interference scheme is the well-known Hong-Ou-Mandel (HOM) interferometer \cite{hong87} where the two fields of modes $c,d$ are superposed at a 50/50 beam splitter (Fig.1(b)). We detect one frequency component, say, $\omega_s$, at one output port and the other ($\omega_i$) in the other port. When coincidence is detected between the two detectors, we register one pair of conjugate photons of $\omega_s,\omega_i$. However, because of the superposition in Eq.(\ref{psi2}), we cannot tell whether $\omega
 _s$ is from $c$ field and $\omega_i$ from $d$ field or vice versa. This indistinguishability in the paths of the photon pair leads to two-photon interference in the same way as the degenerate case in HOM interference effect. A simple calculation will confirm the argument above.

Consider the outputs from the beam splitter in Fig.1(b). They are connected to the $c,d$ fields in Eq.(\ref{Ecd}) by
\begin{eqnarray}
\hat E_1^{(+)}(t)= [\hat E_c^{(+)}(t+\delta\tau)+i\hat E_d^{(+)}(t)]/\sqrt{2},\cr
\hat E_2^{(+)}(t)= [\hat E_d^{(+)}(t)+i\hat E_c^{(+)}(t+\delta\tau)]/\sqrt{2}.\label{E12}
\end{eqnarray}
Here we introduced a delay $\delta\tau$ in the $c$ field. After the filters, we have the field operators at the detectors as
\begin{eqnarray}
\hat E_{D1}^{(+)}(t)= [\hat a_c(\omega_s)e^{-i\omega_s(t+\delta\tau)}+i\hat a_d(\omega_s)e^{-i\omega_st}]/\sqrt{2},\cr
\hat E_{D2}^{(+)}(t)= [\hat a_d(\omega_i)e^{-i\omega_it}+i\hat a_c(\omega_i)e^{-i\omega_i(t+\delta\tau)}]/\sqrt{2}.\label{ED12}
\end{eqnarray}
We may now calculate the probability of two-photon coincidence measurement as
\begin{eqnarray}
P_2(\tau) \propto \langle \hat E_{D2}^{(-)}(t)\hat E_{D1}^{(-)}(t+\tau)\hat E_{D1}^{(+)}(t+\tau)\hat E_{D2}^{(+)}(t)\rangle\label{PDt2}
\end{eqnarray}
and we have after some straightforward manipulation
\begin{eqnarray}
P_2(\tau) \propto 1- \cos(\omega_s-\omega_i)\delta\tau. \label{sbt}
\end{eqnarray}
Note that the detection times $t, t+\tau$ do not appear in Eq.(\ref{sbt}) and the detection probability only depends on the delay $\delta\tau$, which can be precisely controlled with a spatial translator. Thus, there is no need for fast detectors to observe the interference effect: the fast temporal beat is transformed into spatial beating. This idea was first demonstrated by Ou and Mandel and co-workers \cite{ou88b,ou88c}.

In practice, there are always some backgrounds that will degrade the interference effect. Let's take the unentangled mixed state
\begin{eqnarray}
\hat \rho_{un} = (|\omega_s\rangle_c|\omega_i\rangle_d\langle \omega_s|_c\langle\omega_i|_d+ |\omega_i\rangle_c|\omega_s\rangle_d\langle \omega_i|_c\langle\omega_s|_d)/2\cr \label{mix}
\end{eqnarray}
as the background. Then the real state is a mixture of Eq.(\ref{mix}) and Eq.(\ref{psi2}):
\begin{eqnarray}
\hat \rho_{sys} = p|\psi_2\rangle\langle\psi_2|+(1-p)\hat \rho_{un},\label{sys}
\end{eqnarray}
where $p<1$ is the probability of the system in the entangled state. With the above state for the system, we can easily find the probability for two-photon coincidence measurement:
\begin{eqnarray}
P_2 \propto 1- p \cos(\omega_s-\omega_i)\delta\tau. \label{P2}
\end{eqnarray}
Considering the state fidelity is defined as $F=\langle\psi_2|\hat \rho_{sys}|\psi_2\rangle$, we have
\begin{eqnarray}
F=p+(1-p)/2=(1+p)/2, \label{F}
\end{eqnarray}
which is directly related to the visibility of the spatial beating in Eq.(\ref{P2}).

Before going to next topics, let us briefly discuss how we can achieve $\phi =\pi/2$ in order to obtain the frequency entangled state $|\psi_2\rangle$ in Eq.(\ref{psi2}) from a SFL with well matched mode. For the SFL without any optical element, there is no phase difference between the two counter-propagating fields, i.e., $\phi =0$. Under this condition, the SFL acts as a perfect reflector \cite{Mortimore88}. But this will lead to the state $|\psi_1\rangle$ in Eq.(\ref{psi1}). The phase difference is introduced by the FPC inserted in the SFL. It can be easily shown that when $\phi =\pi/2$, the SFL acts as a 50/50 beam splitter, i.e., $I_c=I_d = I_{in}/{2}$, where $I_{c}$ and $I_{d}$ denote the power at the $c$ and $d$ ports, respectively, for an incident pump power $I_{in}$. This provides us a way to set $\phi =\pi/2$ in the experiment.

\section{Polarization mode match}

Because Eq.(\ref{eq7}) results from superposition of different fields, besides the control of the phase difference $\phi $ in Eq.(\ref{eq7}), mode matching between the photon pairs $\left| \omega
_s,\omega _i\right\rangle_a$ and $\left| \omega
_s,\omega _i\right\rangle_b$ is also required to obtain the frequency entangled state in Eq.(\ref{psi2}). Spatial mode matching is automatically satisfied in a single mode fiber.  However, polarization modes need to be carefully matched due to the birefringence inevitably induced by optical fiber and FPC~\cite{Culshaw06}. Since the polarization of photon pairs
generated in our SFL is dominantly parallel to the pump, to understand how to realize the required ideal mode matching, we first
briefly analyze the polarizations of the pumps propagating in the SFL.

Similar to the analysis in Ref.\cite{Mortimore88}, we assume the SFL lies in the $xz$ plane, and $y$ axis
is perpendicular to both the $x$ and $z$ axes, as shown in Fig.2. For simplicity, we assume the birefringence in SFL is only introduced by FPC. Moreover, by introducing three points, $I$, $II$ and $III$, we divide the fiber inside the SFL into two segments, $I-II$ and $II-III$, and FPC is placed in segment $II-III$. In this reference frame, the Jones vector of the incident pump field is written as
\begin{eqnarray}
\overrightarrow{E}_{in}=\left(
\begin{array}{l}
E_x \\
E_y
\end{array}
\right),
\end{eqnarray}
where $E_{x(y)}$ is the complex amplitude component of $\overrightarrow{E}_{in}$ decomposed along $x(y)$ axis. Passing through the 50/50 fiber coupler, $\overrightarrow{E}_{in}$  is equally split into two pump fields $\overrightarrow{E}_{a_{in }}$ and $\overrightarrow{E}_{b_{in}}$, which start to propagate in SFL from clockwise and counter-clockwise directions, respectively. According to the transfer matrix of 50/50 fiber coupler, we have
\begin{eqnarray}
&&\overrightarrow{E}_{a_{in }}=\left(
\begin{array}{l}
E_{ax_{in}} \\
E_{ay_{in}}
\end{array}
\right) =\frac 1{\sqrt{2}} \left(
\begin{array}{l}
E_x \\
E_y
\end{array}
\right)
\label{Eain}
\end{eqnarray}
and
\begin{eqnarray}
&&\overrightarrow{E}_{b_{in}}=\left(
\begin{array}{l}
E_{bx_{in}} \\
E_{by_{in}}
\end{array}
\right) =\frac i{\sqrt{2}} \left(
\begin{array}{l}
E_x \\
E_y
\end{array}
\right)
\label{Ebin}
\end{eqnarray}
where $E_{a(b)x_{in}}$ and $E_{a(b)y_{in}}$ are the complex amplitude components of $\overrightarrow{E}_{a(b)_{in}}$ decomposed along $x$ and $y$ axes, respectively. The $90$-degree phase in $b$-field in Eq.(\ref{Ebin}) is from the cross-coupling in the 50/50 fiber splitter/coupler.

In the SFL, for the light field propagates from point $I$ ($II$) to point $II$ ($I$), the sign of the component decomposed along $x$ axis will be changed, which is equivalent to the light fields passing through a Jones matrix
\begin{eqnarray}
\mathbb{J}_1=\left(
\begin{array}{ll}
-1 & 0 \\
0 & 1
\end{array}
\right).
\end{eqnarray}
For the pump field
transmitting in clockwise and counter-clockwise directions, the Jones matrix of the FPC is $\mathbb{J}_c$ and $\widetilde{\mathbb{J}}_c$, respectively, where $\widetilde{\mathbb{J}}_c$ is the transpose of $\mathbb{J}_c$ and $\mathbb{J}_c$ is given by
\begin{eqnarray}
\mathbb{J}_c=\left(
\begin{array}{ll}
J_{xx} & J_{xy} \\
J_{yx} & J_{yy}
\end{array}
\right).
\end{eqnarray}
Because the energy of light fields passing through FPC from clock- and  anticlock-wise directions is the same, we have $\mathbb{J}_c^{*}\widetilde{\mathbb{J}}_c=\widetilde{\mathbb{J}}_c^{*}\mathbb{J}_c=\mathbb{I}$, where $\mathbb{I}$ is the unit matrix. Based on this relation, it is straightforward to obtain
\begin{eqnarray}
\left| J_{xx}\right| ^2=\left| J_{yy}\right| ^2,
\label{xx=yy}
\end{eqnarray}
\begin{eqnarray}
\left| J_{xy}\right| ^2=\left| J_{yx}\right| ^2
\label{xy=yx},
\end{eqnarray}
and
\begin{eqnarray}
J_{xx}^{*}J_{yx}+J_{xy}^{*}J_{yy}=J_{yx}^{*}J_{xx}+J_{yy}^{*}J_{xy}=0.
\label{dia=0}
\end{eqnarray}

After traveling through the fiber inside the SFL, $\overrightarrow{E}_a$ and $\overrightarrow{E}_b$, the two pumps arrive at the 50/50 coupler and are given by
\begin{eqnarray}
\overrightarrow{E}_a=\left(
\begin{array}{l}
E_{ax} \\
E_{ay}
\end{array}
\right) =\mathbb{J}_c\mathbb{J}_1 \left(
\begin{array}{l}
E_{a_{in}x} \\
E_{a_{in}y}
\end{array}
\right),\label{Ea}
\end{eqnarray}
\begin{eqnarray}
\overrightarrow{E}_b=\left(
\begin{array}{l}
E_{bx} \\
E_{by}
\end{array}
\right) =\mathbb{J}_1\widetilde{\mathbb{J}}_c \left(
\begin{array}{l}
E_{b_{in}x} \\
E_{b_{in}y}
\end{array}
\right),\label{Eb}
\end{eqnarray}
where $E_{a(b)x}$ and $E_{a(b)y}$ are the components of $\overrightarrow{E}_{a(b)}$ decomposed along $x$ and $y$ axes, respectively. After expanding Eqs.(\ref{Ea}) and (\ref{Eb}), we obtain
\begin{eqnarray}
&&E_{ax}=\frac 1{\sqrt{2}}(-J_{xx}E_x+J_{xy}E_y)\begin{array}{ll}
\end{array}\label{Eax}\\
&&E_{ay}=\frac 1{\sqrt{2}}(-J_{yx}E_x+J_{yy}E_y)\begin{array}{ll}
\end{array}\\
&&E_{bx}=\frac i{\sqrt{2}}(-J_{xx}E_x-J_{yx}E_y)\\
&&E_{by}=\frac i{\sqrt{2}}(J_{xy}E_x+J_{yy}E_y)\begin{array}{ll}
\end{array}
\label{Eby}
\end{eqnarray}

After recombination at the 50/50 fiber coupler, the two pumps $\overrightarrow{E}_a$ and $\overrightarrow{E}_b$ then generate two exit fields:
\begin{eqnarray}
&&\overrightarrow{E}_c=\left(
\begin{array}{l}
E_{cx} \\
E_{cy}
\end{array}
\right) =\frac 1{\sqrt{2}}\left(
\begin{array}{c}
iE_{ax}+E_{bx} \\
iE_{ay}+E_{by}
\end{array}
\right) \cr&&~~~~
=i\left(
\begin{array}{c}
-J_{xx}E_x+\frac 12(J_{xy}-J_{yx})E_y \\
J_{yy}E_y+\frac 12(J_{xy}-J_{yx})E_x
\end{array}
\label{Ec}
\right)
\end{eqnarray}
and
\begin{eqnarray}
&&\overrightarrow{E}_d=\left(
\begin{array}{l}
E_{dx} \\
E_{dy}
\end{array}
\right) =\frac 1{\sqrt{2}}\left(
\begin{array}{l}
iE_{bx}+E_{ax} \\
iE_{by}+E_{ay}
\end{array}
\right) \cr&&~~~~
=\frac 12(J_{xy}+J_{yx})\left(
\begin{array}{l}
E_y \\
-E_x
\end{array}
\right)
\label{Ed}
\end{eqnarray}
where $E_{c(d)x}$ and $E_{c(d)y}$ are the components of $\overrightarrow{E}_{c(d)}$ decomposed along $x$ and $y$ axes, respectively.

From Eqs.(\ref{Eax})-(\ref{Eby}), we find that for arbitrary input polarization of $\overrightarrow{E}_{in}$, the polarization modes of the pump fields $\overrightarrow{E}_{a}$ and $\overrightarrow{E}_{b}$ are usually not the same except for some special Jones matrix $\mathbb{J}_c$. In order to have the same polarization for $a$ and $b$ fields, i.e., $E_{ax}/E_{ay} = E_{bx}/E_{by}$, we find that one of the following relations needs to be satisfied:
\begin{eqnarray}
J_{xy} + J_{yx}=0
\label{ed=0}
\end{eqnarray}
\begin{eqnarray}
J_{xx}=0=J_{yy}~ \mathrm{and} ~J_{xy}=J_{yx}
\label{ec=0}
\end{eqnarray}
\begin{eqnarray}
J_{xx}E_x^2-J_{yy}E_y^2= (J_{xy} - J_{yx})E_xE_y
\label{E-match}
\end{eqnarray}
When Eq.(\ref{ed=0}) or Eq.(\ref{ec=0}) is valid, we have $\overrightarrow{E}_d=0$ or $\overrightarrow{E}_c=0$, which means that mode matching is automatically satisfied when the SFL functions as a high reflector or high transmitter~\cite{Fiorentino02,Li05a}.
Otherwise, if $\overrightarrow{E}_{c(d)}\neq 0$, for a specified Jones matrix $\mathbb{J}_c$, the mode matching also depends on the polarization of $\overrightarrow{E}_{in}$, as determined in Eq.(\ref{E-match}).

It is clear from Eqs.(\ref{Ec}, \ref{Ed}) that the polarization mode of transmitted pump field $\overrightarrow{E}_d$ only depends on that of the incident pump $\overrightarrow{E}_{in}$, whereas the polarization mode of the reflected field $\overrightarrow{E}_c$ varies with $\overrightarrow{E}_{in}$ and the Jones matrix $\mathbb{J}_c$.
It should be noted that once the polarizations of the two counter-propagating pumps $\overrightarrow{E}_{a}$ and $\overrightarrow{E}_{b}$ are matched, the polarizations of the reflected and transmitted pump fields $\overrightarrow{E}_c$ and $\overrightarrow{E}_d$ will match as well.

Using Eqs.(\ref{xx=yy})-(\ref{dia=0}) in Eq.(\ref{E-match}) and after some manipulation, we find
\begin{eqnarray}
J_{xx}E_y^{*2}-J_{yy}E_x^{*2}=(J_{yx}-J_{xy})E_x^{*}E_y^{*}.
\label{V-Ein}
\end{eqnarray}
From Eq.(\ref{V-Ein}), we can then infer that once $\overrightarrow{E}_{a}$ and $\overrightarrow{E}_{b}$ are polarization-matched for the pump $\overrightarrow{E}_{in}$, as in Eq.(\ref{E-match}), the polarization mode matching condition will be preserved for the orthogonal input pump
\begin{eqnarray}
\overrightarrow{E}_{in}^{\perp }\equiv \left(
\begin{array}{l}
E_y^{*} \\
-E_x^{*}
\end{array}
\right).\label{Eperp}
\end{eqnarray}

We would like to point out that if the polarization modes of the photon pairs $\left| \omega
_s,\omega _i\right\rangle_a$ and $\left| \omega
_s,\omega _i\right\rangle_b$ are not well matched, the purity and fidelity of the generated frequency entanglement will be degraded. For example, if pump fields $\overrightarrow{E}_{a}$ and $\overrightarrow{E}_{b}$ are orthogonal, the polarizations of the photon pairs $\left| \omega
_s,\omega _i\right\rangle_a$ and $\left| \omega
_s,\omega _i\right\rangle_b$ will also be orthogonal. In this case, signal and idler photons will randomly go to mode $c$ or $d$ even though the SFL functions as a 50/50 power splitter for the incident pump, and the output of the SFL is in a mixed state of $\left| \psi\right\rangle_1$ and $\left| \psi\right\rangle_2$. The existence of $\left| \psi\right\rangle_1$ contributes to the background photons of frequency entanglement $\left| \psi\right\rangle_2$, which will cause the degradation of fidelity $F$.

Combining what we have found above about the phase difference and the polarization control in the SFL, we summarize how to generate the frequency entangled photon pairs $\left| \psi_2\right\rangle =(\left| \omega _s\right\rangle _c\left|
\omega _i\right\rangle _d+\left| \omega _i\right\rangle _c\left| \omega
_s\right\rangle _d)/\sqrt{2}$ as: (i) the two counter-propagated pumps (and photon pairs) have the same polarization when they recombine at the 50/50 coupler or the fields at the output ports $c$ and $d$ have the same polarization; and (ii) the transmission (reflection) efficiency of the SFL is $50\%$. We can carefully adjust the FPC in the SFL to achieve these conditions, as we will see in the next section.

\section{Experiment}

Our experimental setup is shown in
Fig.3. Signal and idler photons at wavelengths of
1544.5\,nm and 1531.9\,nm, respectively, are produced in a
SFL (Fig.3(a)) consisting of a 50/50 fiber coupler (FC1) spliced
to 300 m of DSF with ZDW $\lambda
_0=1538\pm 2$\,nm at 77 K. The DSF is submerged in liquid nitrogen to
reduce the Raman scattering (RS). The linearly polarized pump pulses with a pulse width of $\sim 4$\,ps and a central wavelength of 1538.2 nm,  are spectrally
carved out from a mode-locked femto-second fiber laser (repetition rate $\approx $40 MHz). To achieve the required power, the pump
pulses are amplified by an erbium-doped fiber amplifier.
The pump pulses are further cleaned up with a band-pass filter
F1 having an FWHM of 0.9 nm. A half wave plate (HWP1) is used to adjust the polarization state of the pump. A 95/5 fiber coupler is used
to split $5\%$ of the pump for power monitoring.

To reliably detect the signal and idler photons, an
isolation between the pump and the signal/idler photons in excess of
100\,dB is required, because of the low efficiency of SFWM in DSF. We achieve these by passing the output of SFL through a filter ensemble F2 (Fig.3(b)), or F3 and F4 (Fig.3(c) or (d)). F2 with an FWHM of 0.7 nm is realized by cascading double grating filters (DGFs)~\cite{Li05a}, composed of grating G1 and G2 or G3, with tunable filters (TF) TF1 and TF2 having a central wavelengths of 1544.5\,nm and 1531.9\,nm, respectively (see Fig.3(b)). Mirror M1 in F2 is used to reflect the pump transmitted through SFL for alignment purpose. F3 (F4) with a supper-Gaussian spectrum and an FWHM of about 0.9 nm is realized by cascading two WDM filters, whose central wavelength is 1544.5 (1531.9)\,nm. The signal (idler) photons are counted by single photon
detectors (SPD) operated in the gated Geiger mode. The $2.5$\,ns
gate pulses arrive at a rate of about 1.29 MHz, which is $1/32$ of the
repetition rate of the pump pulses, and the dead time of the gate
is set to be 10 $\mu$s.

\subsection{Polarization mode match and adjustment of phase difference}

For the purpose of polarization mode match, laser pulses at signal wavelength (1544.5 nm), spectrally carved out from the mode-locked fiber laser and path-matched with the pump pulses, are injected into the SFL through a 90/10 fiber coupler (Fig.3(a)). In principle, the required mode matching could be achieved by making the polarization of the reflected pump ($c$-field) the same as that of the transmitted pump ($d$-field). But this is practically not realizable since the two pump fields propagate along different fibers after they come out of the SFL. This difference makes it impossible to compare the polarizations of the two pump fields since different fibers have different birefringence.
To circumvent this, we need the help of an auxiliary beam at the signal wavelength injected into the SFL from $d$ port (see Fig.3(a)). When the polarization of the injected signal field is the same as the input pump field, the gain of FWM inside the SFL is at the maximum. This can be used to match the polarizations of the injected signal at $d$ port and the input pump at $c$ port. Once this is done, the reflected output signal field (coming out from $d$ port) will have the same polarization as the output pump field at $c$ port due to the symmetry of SFL. Thus matching the  polarization of the output signal field at $d$ port with the output pump field at $d$ port is equivalent to matching the polarizations of the pump fields at $c$ and $d$ ports. Since the two fields to be compared are now  all out of the same port ($d$ port), they experience the same birefringence in the same fiber.

So, both the pump and the signal pulses are launched simultaneously into the SFL with HWP1 positioned at 0 degree. The reflected pump, obtained by passing the output of Circ ($c$ output port) through a 1-nm bandwidth TF with a pass-band the same as F1, is monitored by a power meter. The photons out of $d$ port passing through FPC3 and a fiber polarization beam splitter (FPBS1) are directed to F2 (see Fig.3(b)). The signal amplification is first maximized by adjusting FPC2 (for matching the polarizations of the injected signal field with that of the input pump field). Then FPC1 and FPC3 are carefully adjusted, so that not only the reflected pump in mode $c$ is 50$\%$ of its maximum, but also the signal and pump passing through F2 can be maximized and minimized simultaneously, which guarantees that the reflected and optimally amplified signal has the same polarization as the transmitted pump in port $d$. Once the adjustment is completed, the inject
 ed signal is blocked and
further measurement is made on the parametric
fluorescence with only the pump field input.

\subsection{Generation of frequency entangled states}

After we obtain the correct polarization mode match and set the phase difference $\phi$ at $\pi/2$, we need to confirm that what we obtain is $|\psi_2\rangle$ state not $|\psi_1\rangle$ state, before we verify the frequency entanglement. This is done by two-photon coincidence measurement in mode $d$ alone (checking $|\psi_1\rangle$ state) and between mode $c$ and mode $d$ (checking $|\psi_2\rangle$ state).

With the pump power fixed at 0.18 mW and FPBS1 removed, we measure the single counts and coincidence of the signal and idler photons as the polarization of the incident pump is varied by rotating HWP1. We perform two experiments. In the first one, both the signal and idler photons in mode $d$ are directed to F2 for measuring the probability of both signal and idler photons coming out of $d$ port or for checking $|\psi_1\rangle$ state. The result is shown in Fig. 4(a), the coincidence rate varies as the orientation of HWP1 is changed. For HWP1 orientated at 0, 45, 90,135 and 180 degree, respectively, the coincidence is very close to the calculated accidental coincidence obtained from the measured single counts. For HWP1 at 22.5, 67.5, 112.5 and 157.5 degree, respectively, the coincidence rates are about at the maximum. In the second experiment, signal and idler photons coming from ports $c$ and $d$ are directly fed to F3 and F4, respectively, without passing through the polarization elements. This is for checking $|\psi_2\rangle$ state. The result is shown in Fig.4(b). The coincidence rate also varies as the orientation of HWP1 changed, but the direction of change is the opposite of Fig.4(a).

We noticed that the rotation of HWP1 does not affect the transmission efficiency of 50$\%$ for the SFL. This is consistent with Eq.(\ref{Ed}), showing the transmission only depends on the Jones Matrix $\mathbb{J}_c$ in SFL. For HWP1 orientated at 0, 90, and 180 degree, respectively, the polarization of the incident pump is actually the same and the well matched mode does not change. For HWP1 orientated at 45 and 135 degree, respectively, the polarization of the incident pump is rotated to its orthogonal mode, the mode matching condition is still preserved, confirming Eq.(\ref{Eperp}). In these situations, the output of the SFL is $\left| \psi_2\right\rangle =(\left| \omega _s\right\rangle _c\left|
\omega _i\right\rangle _d+\left| \omega _i\right\rangle _c\left| \omega
_s\right\rangle _d)/\sqrt{2}$, but not $\left| \psi_1\right\rangle =(\left| \omega _s,\omega _i\right\rangle
_c-\left| \omega _s,\omega _i\right\rangle _d)/\sqrt{2}$. So the coincidence rate of signal and idler photons in mode $d$ is expected to be close to the accidental coincidence rate, whereas the coincidence rate of signal and idler in mode $c$ and $d$, respectively, is the highest. One sees that in Fig.4(a), the coincidence for HWP1 at 45 and 135 degree, respectively, is not at the same height as that for HWP1 at 0 degree, we think this discrepancy may be originated from the imperfection in HWP1.

The results in Fig.4 agree with our analysis that non-ideal polarization mode matching of photon pairs will degrade the purity of frequency entanglement photon pairs. Notice that in Fig.4(b), the maximum of the true coincidence rate (HWP1 positioned at 0, 45, 90,135 and 180 degree), which is the difference between the coincidence and the accidental coincidence  is twice of the minimum (HWP1 positioned at 22.5, 67.5, 112.5 and 157.5 degree). Comparing Fig.4(a) with Fig.4(b), and taking into account the difference in efficiency and bandwidth of F2, F3, and F4, we find the value of maximum true coincidence rate in Fig.4(a) is the same as that of the minimum coincidence rate in Fig.4(b). HWP1 is all positioned at 22.5, 67.5, 112.5 and 157.5 degree for these cases. All above can be understood as follows.

For HWP1 at 22.5 and 67.5 (112.5 and 157.5) degree, respectively, the incident pump can be equally decomposed along the two orthogonal modes: one mode corresponds to, say, $\overrightarrow{x}$; the other corresponds to  $\overrightarrow{y}$. In this reference frame, when the two counter-propagating pumps meet at FC1, if the phase difference of the pumps along $\overrightarrow{x}$ is  $\phi_x =\pi/2 $, then we have  $\phi_y =-\pi/2 $ for the pumps decomposed along $\overrightarrow{y}$ \cite{com}. This means that if the two counter-propagating pumps are originally polarized along 45 degree (HWP1 at 22.5 degree), they will be orthogonally polarized when they meet at FC1, and so are the photon pairs. Therefore, the signal and idler photons created in pairs randomly go to mode $c$ or $d$, and the output of SFL is a mix of $\left| \psi_1\right\rangle$ and $\left| \psi_2\right\rangle$ with equal amplitude. As a result, the true coincidence rate of signal and idler photons is about half of the achievable highest rate, as observed in both Fig.4(a) and Fig.4(b).

To further improve the purity of the frequency entangled photon pairs, FPC3 and FPBS1 are used to select the scattered photons co-polarized with the pump, and so are FPC4 and FPBS2 (see Fig.3(c)). This is because in the process of generating photon pairs via SFWM in DSF, the accompanying Raman scattering background photons which include co- and cross-polarized photons would inevitably degrade the purity of the entangled photon pairs, and the cross-polarized photons can be suppressed by using a polarization beam splitter~\cite{Li04}. To demonstrate the improved purity, similar to previous experiment, we fix the HWP1 at 0 degree and measure the single counts and coincidence of the co-polarized signal and idler photons as a function of pump power in the two cases. When both the co-polarized signal and idler photons in mode $d$ are directed to F2, no obvious difference between the measured coincidence and calculated accidental coincidence rates are observed, as shown in Fig.5(a).
When the co-polarized signal and idler photons come from modes $c$ and $d$ are fed to F3 and F4, respectively, the coincidence rate is much higher than that of the calculated accidental coincidence, as shown in Fig.5(b). Comparing the ratio between the rates of coincidence and accidental coincidence in Fig.5(b) with that in Fig.4(b), for the average pump power of 0.18 mW and the orientation of HWP1 at 0 degree, we find the ratio in Fig.5(b) is about 16, whereas the ratio in Fig.4(b) is about 13. This shows the improvement in the purity of the entangled photon pairs. Moreover, this set of data also confirms the output state of SFL is $\left| \psi_2\right\rangle =(\left| \omega _s\right\rangle _c\left|
\omega _i\right\rangle _d+\left| \omega _i\right\rangle _c\left| \omega
_s\right\rangle _d)/\sqrt{2}$, in which if the signal photon appears in mode $c$ ($d$), its twinborn idler photon will be present in mode $d$ ($c$).  We are now ready to measure the frequency entanglement.

\subsection{Verification of frequency entanglement}

The frequency entanglement can be confirmed by the spatial beating effect~\cite{ou88b}. Signal and idler photons with non-degenerate frequencies in modes $c$ and $d$ are carefully path matched and simultaneously directed to a $50/50$ fiber coupler (FC2) from two input ports (see Fig.3(d)), respectively.
Before coupling in FC2, co-polarized signal and idler photons in mode $c$ are delayed by reflector
mirrors mounted on a translation stage. To ensure the two input fields of FC2 have
the identical polarization, the polarization of the delayed photons is properly adjusted by FPC5.
The two outputs of FC2 are passed through filters F3 and F4, and detected by SPD1 and SPD2, respectively. For the generated frequency entangled state $\left| \psi_2\right\rangle =(\left| \omega _s\right\rangle _c\left|
\omega _i\right\rangle _d+\left| \omega _i\right\rangle _d\left| \omega
_s\right\rangle _d)/\sqrt{2}$, the coincidence counting probability between SPD1 and SPD2 is given by Eq.(\ref{sbt}) or Eq.(\ref{P2}) if the state is not pure. However, in practice, the signal and idler fields have a broadband. A detailed multimode theory was given in Ref.\cite{ou88b} and the coincidence probability is proportional to
\begin{eqnarray}
P_2\propto 1-V f(\delta \tau )\cos [(\omega _i-\omega _s) \delta \tau ] ,
\label{multi}
\end{eqnarray}
where $V$ is the visibility, $\delta \tau$ is the optical time delay between the two paths
from the 50/50 coupler FC1 to the 50/50 coupler FC2. The function $f(\delta \tau )$ is associated with the spectra of the detected signal and idler fields and is usually determined by the filters F3 and F4 because of the broad bandwidth of the generated signal and idler fields. In our experiment, the frequency difference $(\omega _i-\omega _s)/2\pi$ equals $1.58\times 10^{12}$\,Hz;  the time delay is associated with the difference in readings of the translation stage $\delta  l$ through the relation $ \delta \tau=2\delta  l/c$, where $c$ is the speed of light in vacuum;
the spectrum of F3(F4) is super-Gaussian shaped and can be approximated as a square function, so $f(\delta \tau )$ is a sinc-function: $f(\delta \tau) \approx \mathrm{sinc} (\sigma \delta \tau)$, where $\sigma $ is determined by the bandwidth of F3(F4).

The observation of frequency entanglement is achieved by measuring the coincidences between the two SPDs as the position of the translation stage is varied (see Fig.3(d)). In the experiment, the average power of pump is about 0.1 mW, and both the single counts and coincidences of SPDs are recorded by the counting system. The photon counting measurement shows that the single counts of signal and idler photons, recorded by SPD1 and SPD2, respectively, stay constant. The deduced production rate of detected photons in signal (idler) band is about 0.02 photons/pulse, from which we find the photon-pair production rate is about 0.013 pairs/pulse~\cite{li08a}. However, the coincidences exhibit an interference pattern in the form of spatial beating (dark
counts of SPDs have been subtracted), as shown in Fig.6. The periodicity of the spatial beating is $\Delta l=0.095$ mm, which corresponds almost exactly to the period $\Delta \tau =2\pi /(\omega _i-\omega _s)=2\Delta l/c$. The solid curve in Fig.6 is the plot of the function $P_{2}$ given by Eq.(\ref{multi}) with the proportional constant, the origin of position, the bandwidth of F3 and F4, and the visibility adjusted for best fit. The result of the fit gives a visibility of $V=(95\pm 2)$\%. One sees that the experimental data reasonably agrees with the fit. There is slight displacement between the data points and the fitted curve. We believe this is caused by the crudeness of the translation stage.

It is worth noting that the visibility $(95\pm 2)$\% is obtained in our experiment without the subtraction of the accidental coincidences, which are from multi-pair contribution. Gisin and co-workers considered the influence of multiple pairs in two-photon interference \cite{Marcikic02pra}. Following the same argument but using a thermal statistics for the multi-pair statistics, we obtain
\begin{eqnarray}
V_{th} = 1 - 2P_p,
\label{multipair}
\end{eqnarray}
where $P_p$ is the probability of single pair in one pulse. With a photon pair production rate of $P_p \sim $ 0.013 pairs/pulse, we have the theoretical upper limit $V_{th} = 0.974$. The discrepancy with the experimental value is mainly due to the existence of co-polarized Raman scattering, which can be decreased by cooling the DSF further~\cite{Nam08}.

Since the visibility of the interference is not 100\%, the state fidelity is not unity. From Eq.(\ref{F}), we find the state fidelity of the system is $F=(1+V)/2= 0.975$.

\section{Summary and Discussion}
In conclusion, we have developed and characterized an all fiber source of frequency entangled photon pairs
at telecom band. Purity of the entangled state is increased by matching the polarization mode of photon pairs counter-propagated in SFL. This mode matching method will be useful for SFL-based quantum-state engineering~\cite{ChenNJP08}. Moreover, the wavelength of this kind of entanglement can be extended to various ranges by using photonic crystal fibers~\cite{Fan05a,Alibart06}. The basic states of the entanglement at frequencies $\left| \omega_{s}\right\rangle$ and $\left| \omega_{i}\right\rangle$ can be used as qubits for QIP, which resembles the schemes in atomic physics where different energy levels are used to realize a qubit. Furthermore, in contrast to the frequency entanglement created by $\chi ^{(2)}$ nonlinear crystals~\cite{ou88b,Seok08,Heonoh03}, our source has the advantage of mode purity. The spatial mode of all the signal and idler photons with frequencies within the gain bandwidth of SFWM is the guided transverse mode of the fiber. Additionally, comparing
 with other kinds of entangled photon pairs in finite dimensions, such as polarization entanglement and time-bin entanglement etc, the state space of our source is an infinite-dimensional continuous variable (frequency). Therefore, we believe that the fiber source of frequency entangled photon pairs
will prove to be useful for developing quantum information technologies~\cite{Mazurenko99}.

Furthermore, this source can be used in recently discovered quantum optical coherent tomography \cite{ser}. The narrow feature at zero delay is much sharper than the Hong-Ou-Mandel dip in the original schemes \cite{ser,ser2} and can achieve better accuracy \cite{omopn89}.

\begin{acknowledgments}
This work was supported in part by NECT-060238,
the NSF of China (No. 60578024, No 10774111), Foundation for Key Project of
Ministry of Education of China(No. 107027), the State Key
Development Program for Basic Research of China (No. 2003CB314904), the Specialized Research Fund for the Doctoral Program of Higher Education of China
(No.20070056084), National High Technology Research and Development Program of
China (No.2007AA03Z447), and 111 Project B0704.

\end{acknowledgments}



\begin{thebibliography}{19}
\bibitem{com}
It can be easily shown following the formalism in Ref.\cite{Mortimore88} that there is a $\pi$-phase shift between the two orthogonal polarizations for the phase difference in two counterpropagating fields in the SFL, that is, $\phi-\phi^{\perp} =\pi$.
\end{thebibliography}

\newpage

\begin{figure}
\includegraphics[width=7cm]{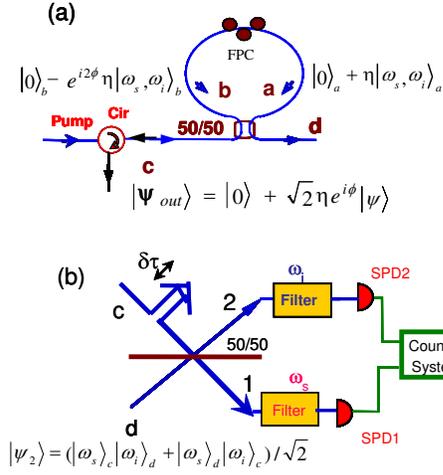}
\caption{(Color online) The conceptual schematic of (a) generating frequency entanglement
by using a Saganc fiber loop (SFL) with well matched mode, and (b) observing quantum interference with frequency entangled two-photon state. Cir, circulator; FPC, fiber polarization controller; SPD, single photon detector.
}
\end{figure}

\begin{figure}
\includegraphics[width=7cm]{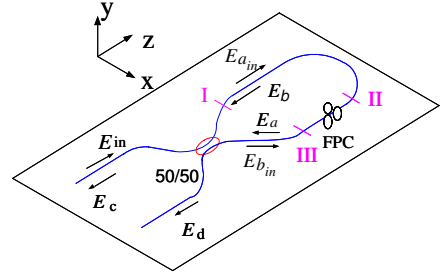}
\caption{(Color online) Polarization reference frame of the pump fields propagating around the SFL. }
\end{figure}

\begin{figure}
\includegraphics[width=8cm]{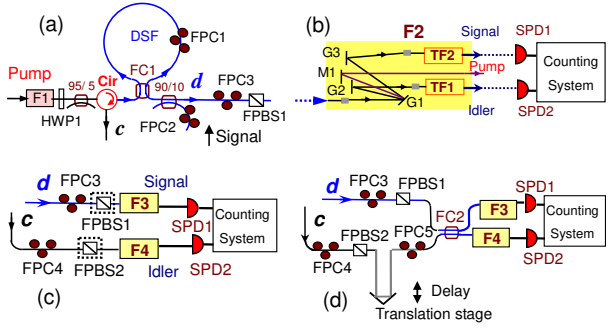}
\caption{(Color online) Experimental setup: (a) Sagnac fiber loop (SFL); (b)  arrangement for polarization mode match and for two-photon
coincidence measurement in output port $d$ for checking
$\left| \psi_1\right\rangle$ state; (c) two-photon coincidence measurement between
output ports $c$ and $d$ for checking $\left| \psi_2\right\rangle$ state; (d)
schematic for verifying frequency entanglement generated from SFL. F, filter; HWP, half wave plate; Cir, circulator; FPC, fiber
polarization controller; FC, 50/50 fiber coupler; FPBS, fiber polarization beam splitter; G, grating; M, mirror; TF, tunable filter; SPD, single photon detector.
}
\end{figure}

\begin{figure}
\includegraphics[width=8cm]{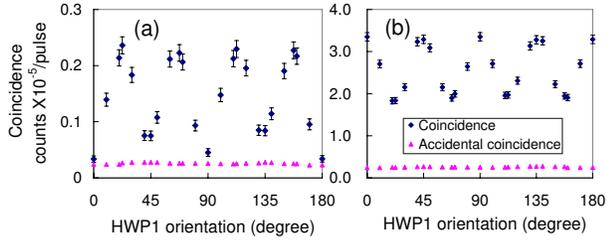}
\caption{(Color online) Coincidence rate (diamonds) and accidental coincidence rate (triangles) of signal and idler photons (a) both in mode $d$ and (b) in modes $c$ and $d$, respectively, as the polarization of the linearly polarized incident pump is varied by rotating HWP1. }
\end{figure}

\begin{figure}
\includegraphics[width=8cm]{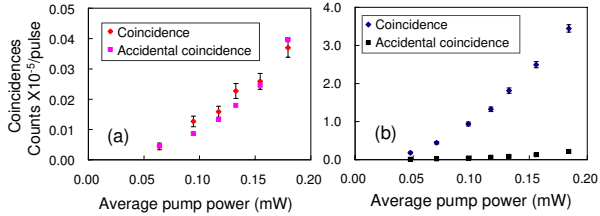}
\caption{(Color online) Coincidence rate of co-polarized signal and idler photons (a) both in mode $d$ and (b) in modes $c$ and $d$, respectively, as the power of pump is varied while keeping HWP1 oriented at 0 degree to ensure well matched mode in SFL.}
\end{figure}

\begin{figure}[htb]
\includegraphics[width=8cm]{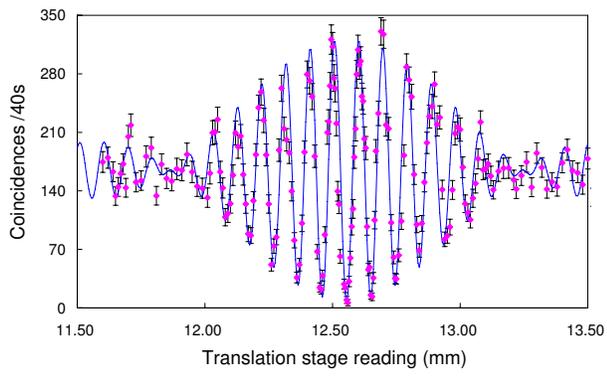}
\caption{(Color online) Coincidence rate versus the position of
the translation stage. The solid curve is a fit to Eq.(\ref{multi}), where $\sigma =2\pi \times 1.09\times 10^{11}$ rad/s, $V =95$\%.}
\end{figure}

\end{document}